\documentstyle[epsfig]{mn}

\newif\ifAMStwofonts

\def\etal{et al. }

\def\aj{Astron.\ J.}
\def\apj{ApJ}
\def\apjl{ApJ\ Lett.}
\def\apjs{ApJ\ Suppl.}

\def\aa{A\&A}
\def\aal{A\&A Lett.}

\def\pasj{{\it PASJ}}

\def\submit{submitted}

\def\mjysr{MJy/sr }
\def\inu{{I_{\nu}}}

\def\bnu{{B_{\nu}}}
\def\msol{{M$_{\odot}$}}
\def\mic{$\mu$m}
\def\cm2{$cm^{-2}$}

\title{The complete submillimetre spectrum of NGC 891}
\author[X. Dupac \etal]
{X. Dupac$^{1,2}$, C. del Burgo$^1$, J.-P. Bernard$^2$, M. Giard$^2$\\
{\Large J.-M. Lamarre$^3$, R.J. Laureijs$^1$, F. Pajot$^4$, I. Ristorcelli$^2$, G. Serra$^2$, J. Tauber$^1$, J.-P. Torre$^5$}\\
1: European Space Agency - ESTEC, Astrophysics Division, Keplerlaan 1, 2201 AZ Noordwijk, The Netherlands\\
2: Centre d'\'Etude Spatiale des Rayonnements (CESR), 9 av. du Colonel Roche, BP4346, 31028 Toulouse cedex 4, France\\
3: LERMA, Observatoire de Paris, 61 av. de l'Observatoire, 75014 Paris, France\\
4: Institut d'Astrophysique Spatiale, Campus d'Orsay, b\^at. 121, 15 rue Cl\'emenceau, 91405 Orsay cedex, France\\
5: Service d'A\'eronomie du CNRS, BP3, F-91371 Verri\`eres-le-Buisson cedex, France}

\date{Accepted
      Received
      in original form}

\pagerange{ -- }
\pubyear{2003}

\begin{document}

\maketitle

\begin{abstract}
Submillimetre maps of NGC 891 have been obtained with the PRONAOS balloon-borne telescope and with the ISOPHOT instrument on board the ISO satellite.
In this article, we also gather data from IRAS and SCUBA to present the complete submillimetre spectrum of this nearby edge-on spiral galaxy.
We derive submillimetre emission profiles along the major axis.
The modified blackbody fits, assuming a single dust component, lead to temperatures of 19-24 K toward the centre and 18-20 K toward the edges, with possible variations of the dust spectral index from 1.4 to 2.
The two-component fits lead to a warm component temperature of 29 K all along the galaxy with a cold component at 16 K.
The interstellar medium masses derived by these two methods are quite different: 4.6$\times$10$^{9}$ M$_{\odot}$ in the case of the one-component model and 12$\times$10$^{9}$ M$_{\odot}$ in the case of the two-component one.
This two-component fit indicates that the cold dust to warm dust ratio is 20 to 40, the highest values being in the wings of this galaxy.
Comparing to dust mass estimates, both estimations of the ISM mass are consistent with a gas to dust mass ratio of 240, which is close to the Milky Way value.
Our results illustrate the importance of accurate submillimetre spectra to derive masses of the interstellar medium in galaxies.

\end{abstract}


\section{Introduction}

NGC 891 is one of the most perfect edge-on \cite{sofue93} spiral galaxies in our neighbourhood.
This peculiarity, and its similarity with the Milky Way galaxy in Hubble type (Sb), optical luminosity and rotational velocity \cite{vanderkruit84}, has allowed it to be intensively studied in order to understand the distribution and the characteristics of the galactic interstellar medium.
It has been widely studied in radio continuum ({\it e.g.} Allen \etal 1978, Dahlem \etal 1994), carbon monoxide (Garc\'\i a-Burillo \etal 1992, Scoville \etal 1993, Sofue \& Nakai 1993, Garc\'\i a-Burillo \& Gu\'elin 1995, Sakamoto \etal 1997), atomic hydrogen (for instance Swaters \etal 1997), molecular hydrogen \cite{valentijn99} and optical extinction \cite{howk97}.
The dust far-infrared continuum emission has been observed by the IRAS satellite (Wainscoat \etal 1997, Rice \etal 1988), the 1.3 mm emission by Gu\'elin \etal (1993), and the submillimetre emission has recently been mapped with SCUBA at 450 and 850 \mic~by Alton \etal (1998) and Israel \etal (1999).
For instance, Alton \etal (1998) showed that the dust submillimetre emission could be fitted by a two-dust component model, assuming a spectral index equal to 2.
However, this was a fit with zero degree of freedom, thus it needs to be confirmed by including other submillimetre measurements.
The accurate knowledge of dust properties in this galaxy, and therefore in other spiral galaxies such as the Milky Way, needs a better submillimetre spectral coverage, in order to properly derive the dust temperature(s) and spectral index.
This is crucial to derive accurate masses of the interstellar medium in galaxies.

In this article, we present new results obtained in the submillimetre range with the ISOPHOT \cite{lemke96} instrument on board of ESA's Infrared Space Observatory \cite{kessler96}, and with the PRONAOS (PROgramme NAtional d'Observations Submillim\'etriques) balloon-borne telescope.
Section 2 presents the observations and data processing, Section 3 presents the results obtained, and Section 4 presents an analysis of the data set.

\section{Observations and data processing\label{obs}}

\subsection{PRONAOS data}

PRONAOS is a French
balloon-borne submillimetre experiment.
Four bolometers cooled at 0.3 K measure the submillimetre flux with sensitivity to low
brightness gradients of about 4 \mjysr in band 1 (200 \mic) and 0.8 \mjysr in band 4 (580 \mic).
The effective wavelengths are 200, 260, 360 and 580 \mic, and the angular resolutions are
2$'$ in bands 1 and 2, 2.5$'$ in band 3 and 3.5$'$ in band 4.
Details about the instrument can be found in Ristorcelli \etal (1998) and Lamarre \etal (1994).
The data which we analyze here were obtained during the second flight of PRONAOS in September 1996, at Fort Sumner, New Mexico.
The data processing and the map-making method, including deconvolution from chopped data, are described in Dupac \etal (2001).
However, for these NGC 891 data, we use a non prior deconvolution method, implemented the same way as the Wiener filter in Dupac \etal (2001).
Indeed, these data are quite noisy, so the optimal Wiener filter may somehow modify the signal, which is not the case with the non prior map-making method.
Then we carefully subtract the background around the galaxy to get the corrected surface brightness, and we smooth the maps of the first three bands with adequate profiles in order to obtain the same angular resolution as the fourth band (3.5$'$).
After having done this, we check the correlation of the pixels in each band with respect to all the other bands in order to detect possible residual offsets between bands.
This happened in band 2 that we rescaled to match the other bands.
The noise fluctuations in the final maps define the spatial relative uncertainty which is 5 MJy/sr at 200 \mic, 4 MJy/sr at 260 \mic~and 2 MJy/sr at 360 and 580 \mic.
The intercalibration error between bands is 5 \% (1 $\sigma$), and the absolute calibration uncertainty is 8 \%.

\subsection{ISOPHOT data}

NGC 891 has been mapped with ISOPHOT using the Astronomical Observation
Template PHT32 in chopped mapping mode with the C200 array detector
(2$\times$2 pixels, 92$''$ per pixel) at the reference wavelengths
170 and 200 \mic~(see details in Laureijs \etal 2001).
The data have been processed using the ISOPHOT Interactive Analysis software PIA v9.1 \cite{gabriel97}.
The data reduction procedure includes ramp linearisation, ramp deglitching, reset interval correction, dark current
subtraction, signal linearisation and signal deglitching.
Each observation has been bracketted with two measurements of the fine calibration source (FCS).
For the flux calibration, we have used the second FCS.
The first quartile normalisation flat-fielding method has been used as
implemented in PIA in order to correct for the remaining responsivity differences of the individual detector pixels.
We smooth the derived 1.5$'$-resolution maps to 3.5$'$ in order to obtain a consistent data set.
The absolute photometric error of ISOPHOT is around 20 \%.

\subsection{IRAS and SCUBA data}

We use the SCUBA 450 and 850 \mic~flux profiles, as well as the high resolution (HiRes) 60 and 100 \mic~flux profiles from the InfraRed Astronomical Satellite survey, as published in Alton \etal (1998).
We smooth the original profiles to the angular resolution of the fourth band of PRONAOS (3.5$'$).

\section{Results\label{res}}

\begin{figure}
\includegraphics[scale=.5]{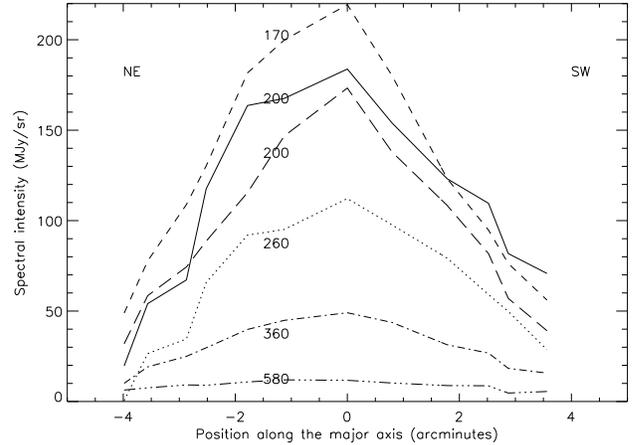}
\caption[]{Major axis intensity profiles of NGC 891 obtained with ISOPHOT and PRONAOS.
The 170 and 200 \mic~ISOPHOT profiles are respectively presented as a dashed line and a long-dashed line.
The angular resolution of these profiles is 1.5$'$.
The 200, 260, 360 and 580 \mic~PRONAOS profiles are respectively plotted as a full line, a dotted line, a dash-dotted line and a dash-dot-dotted line.
The angular resolution is 2$'$ in the first two bands, 2.5$'$ in the third and 3.5$'$ in the fourth.
}
\label{ncprofiles}
\end{figure}

The information obtained is presented as intensity profiles along the major axis in Fig. \ref{ncprofiles}.
We have checked the photometry of all 3.5$'$-resolution maps with respect to all the others, by plotting pixel-pixel diagrams.
No noise offset between bands could be detected.
In particular, both 200 \mic~bands (ISOPHOT and PRONAOS) are in good agreement with each other, which gives a good global consistency to the data set.

All bands of PRONAOS and ISOPHOT show a good symmetry between both sides of the galaxy, which extends approximatively 4$'$ away from the centre.
This extension is similar to the one of the SCUBA maps of Alton \etal (1998) and Israel \etal (1999).
Fig. \ref{ncprofiles} shows a large enhancement of the intensities at an angular distance of about 1.5$'$ away from the galactic centre along the north-eastern side of the galaxy.
This is clearly visible in the 200 and 260 \mic~PRONAOS bands which have the adequate angular resolution (2$'$), as well as in both ISOPHOT bands.
This enhancement can also be seen (at the same place) in the SCUBA maps presented by Alton \etal (1998) and Israel \etal (1999).
Another enhancement, though less large, is visible at the same angular distance on the other side of the galaxy on SCUBA maps.
The explanation for these enhancements could be spiral arms or a molecular ring encircling the galactic bulge.

\section{Analysis\label{ana}}

\subsection{The dust submillimetre emission}

\begin{figure}
\includegraphics[width=9cm,clip=true,viewport=10 10 500 350]{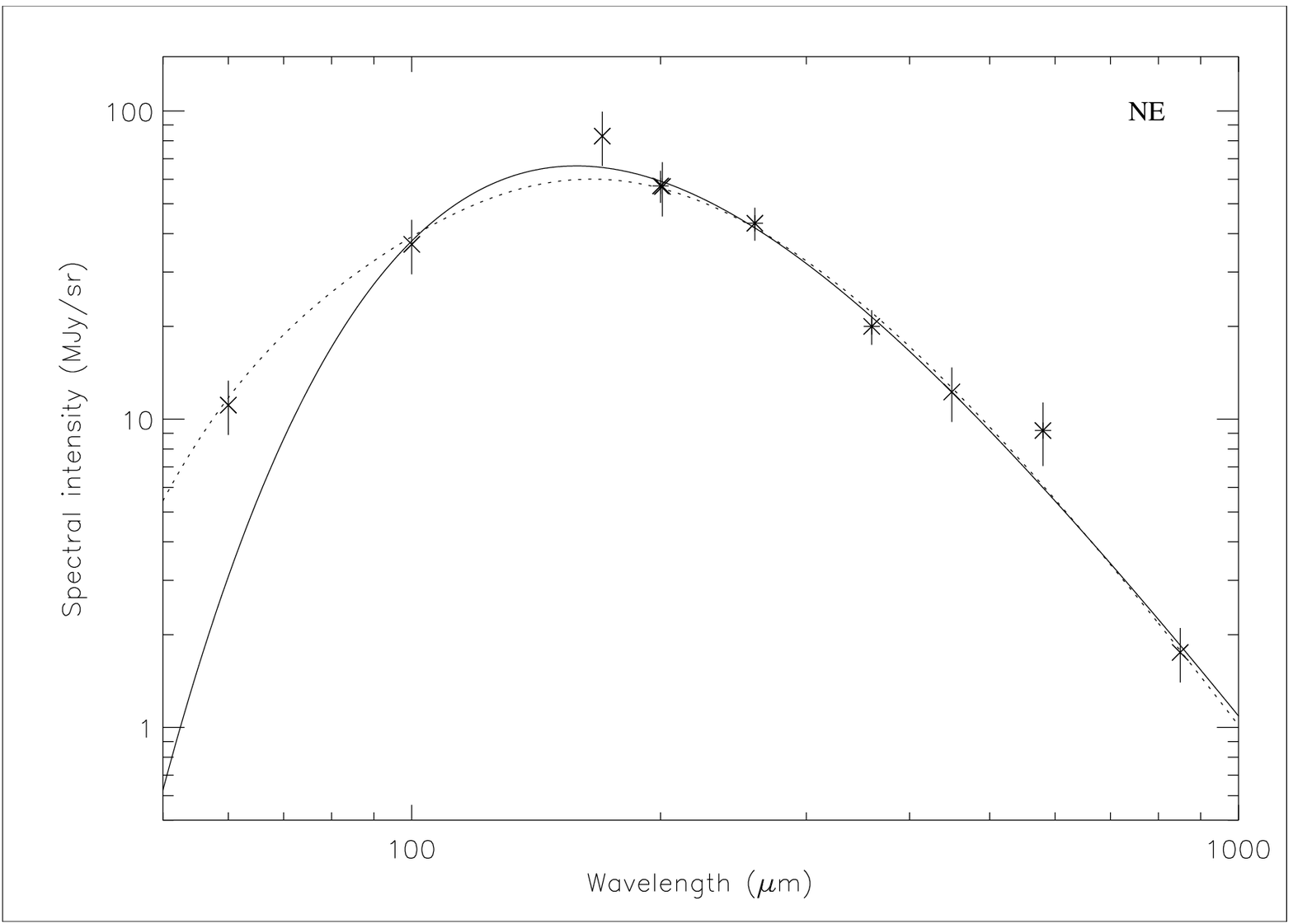}
\includegraphics[width=9cm,clip=true,viewport=10 10 500 350]{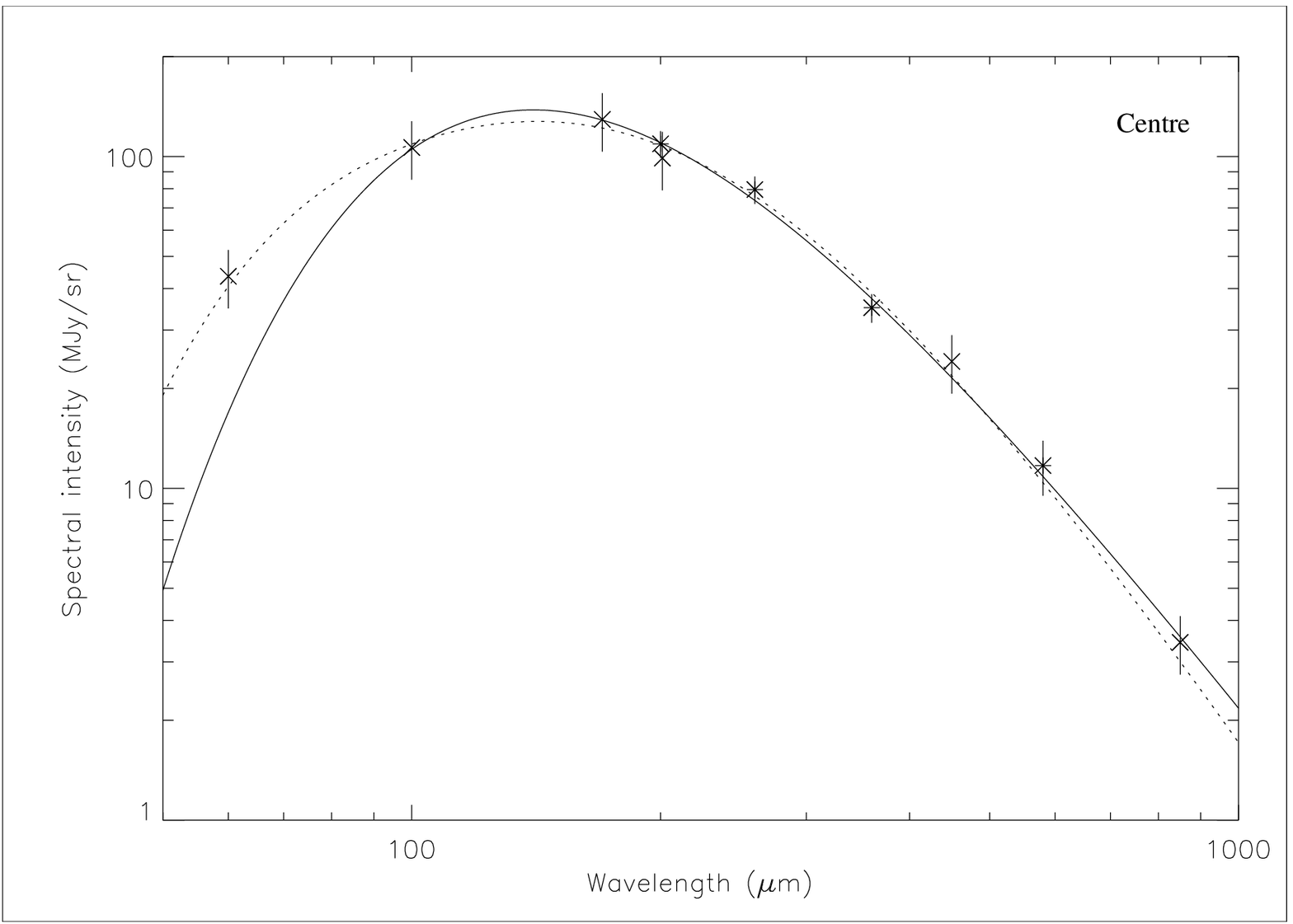}
\includegraphics[width=9cm,clip=true,viewport=10 10 500 350]{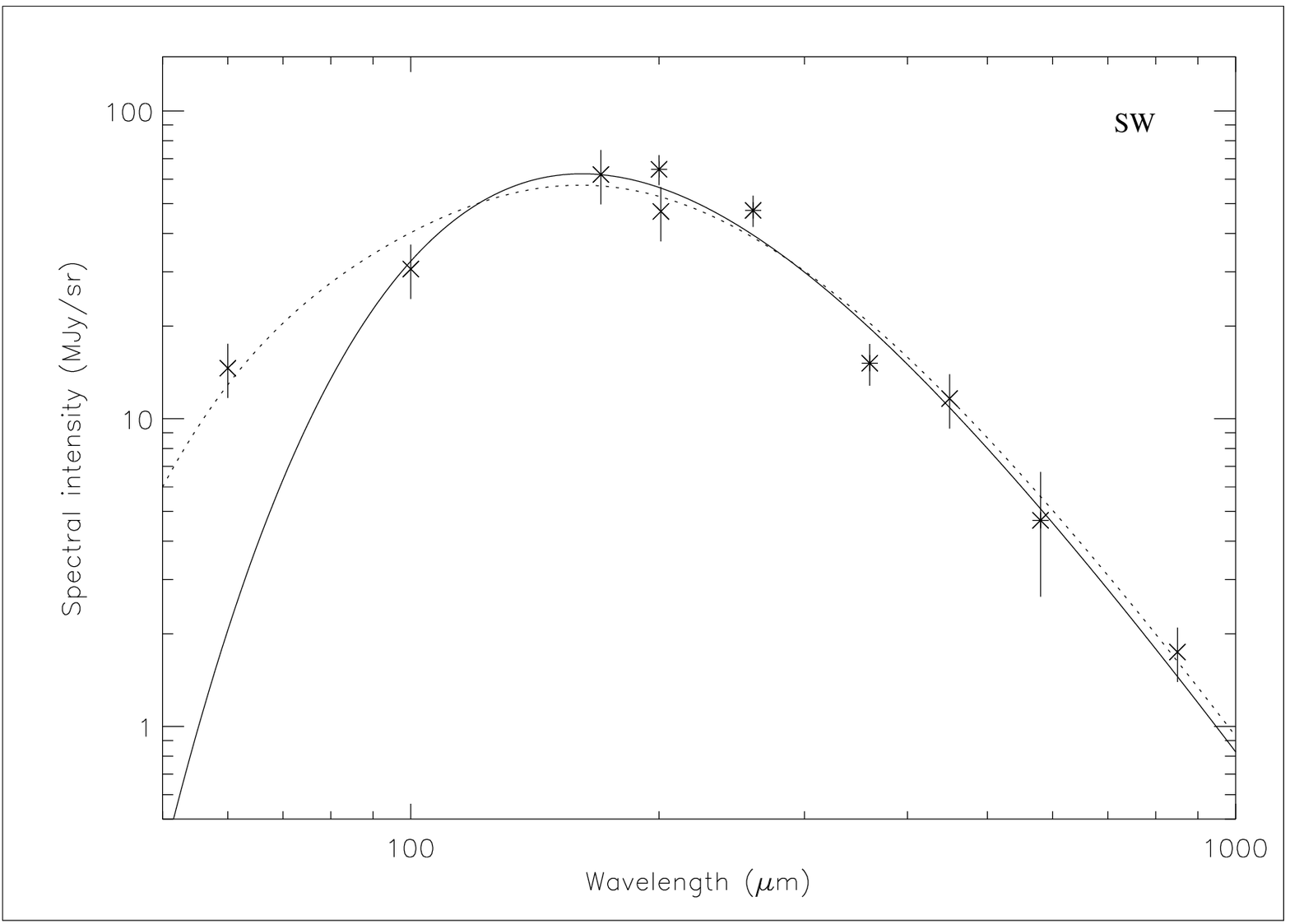}
\caption[]{Spectra of the NGC 891 galaxy.
The effective wavelengths are 200, 260, 360 and 580 \mic~(PRONAOS, asterisks), 170 and 200 \mic~(ISOPHOT, crosses), 100 \mic~(IRAS/HiRes, crosses), and 450 and 850 \mic~(SCUBA, crosses).
The full lines are the results of the one-component fits with $\beta$ free, without the IRAS/HiRes 60 \mic~points.
The dotted lines are the results of the two-component fits ($\beta$=2) including the 60 \mic~points.
}
\label{spec}
\end{figure}

We present in Fig. \ref{spec} the spectra obtained towards the galactic centre and 3$'$ away from the centre on either side.
The plotted error bars are dominated by the global calibration uncertainty of each instrument.
As can be seen in Fig. \ref{spec}, the submillimetre spectral energy density is very well constrained with the present data.

Several fitting procedures have been applied to the three spectra displayed in Fig. \ref{spec}.
The usual simple modelization of the large-grain emission is the modified blackbody, which obeys the following equation:

\begin{equation}
\inu = \epsilon_0 \; \bnu(\lambda,T) \; ({\lambda \over \lambda_0})^{-\beta}
\end{equation}

where $\lambda$ is the wavelength, $\epsilon_0$ the emissivity of the observed dust column density at $\lambda_0$, T the temperature of the grains, $\beta$ the spectral
index and $\bnu$ the Planck function.
We fit the data using a one-component modified blackbody model, with and without the 60 \mic~point (which may be contaminated by very small grain emission, see D\'esert \etal 1990) and assuming $\beta$=2 or not.
We also fit the data using a two-component model, assuming $\beta$=2 for both components, which have different temperatures.
A summary of the results is shown in Table \ref{fits}.

\begin{table}
\caption[]{\label{fits}
Results of the fits toward the centre of the galaxy and 3$'$ away along the major axis on either side (north-east and south-west), for the different methods investigated.
d.o.f. stands for degrees of freedom.
}
\begin{center}
\begin{tabular}{l}
\hline
One component, $\beta$ free, without 60 \mic \\
\end{tabular}
\begin{tabular}{llll}
 & T (K) & $\beta$ & $\chi^2$/d.o.f.\\
Centre & 23.5 $\pm$ 2.6 & 1.41 $\pm$ 0.23 & 0.31/6\\
N-E & 19.6 $\pm$ 1.8 & 1.70 $\pm$ 0.25 & 0.66/6 \\
S-W & 18.1 $\pm$ 1.6 & 1.96 $\pm$ 0.29& 1.51/6\\
\end{tabular}
\begin{tabular}{l}
\hline
One component, $\beta$ = 2, without 60 \mic \\
\end{tabular}
\begin{tabular}{llll}
 & T (K) & $\beta$ & $\chi^2$/d.o.f.\\
Centre & 18.8 $\pm$ 0.8 & 2 & 0.97/7 \\
N-E & 17.8 $\pm$ 0.7 & 2 & 0.75/7 \\
S-W & 17.9 $\pm$ 0.6 & 2 & 1.30/7\\
\hline
\end{tabular}
\begin{tabular}{l}
Two components, $\beta$ = 2, without 60 \mic \\
\end{tabular}
\begin{tabular}{llll}
 & T$_{warm}$ (K) & T$_{cold}$ (K) & $\chi^2$/d.o.f.\\
Centre & 21.0$^{+3.5}_{-3.6}$ & 9.0$^{+13.8}_{-0.0}$ & 1.94/5 \\
N-E & 18.3$^{+2.0}_{-4.3}$ & 7.7$^{+13.1}_{-0.0}$ & 3.86/5 \\
S-W & 18.3$^{+1.8}_{-3.0}$ & 5.0$^{+14.8}_{-0.0}$ & 8.43/5\\
\hline
\end{tabular}
\begin{tabular}{l}
One component, $\beta$ free, with 60 \mic \\
\end{tabular}
\begin{tabular}{llll}
 & T (K) & $\beta$ & $\chi^2$/d.o.f.\\
Centre & 27.8 $\pm$ 1.8 & 1.15 $\pm$ 0.16 & 0.74/7 \\
N-E & 23.4 $\pm$ 1.4 & 1.34 $\pm$ 0.18 & 1.51/7 \\
S-W & 22.5 $\pm$ 1.7 & 1.44 $\pm$ 0.22 & 3.39/7\\
\hline
\end{tabular}
\begin{tabular}{l}
One component, $\beta$ = 2, with 60 \mic \\
\end{tabular}
\begin{tabular}{llll}
 & T (K) & $\beta$ & $\chi^2$/d.o.f.\\
Centre & 20.6 $\pm$ 0.7 & 2 & 2.83/8 \\
N-E & 19.0 $\pm$ 0.6 & 2 & 2.42/8 \\
S-W & 18.6 $\pm$ 0.6 & 2 & 3.33/8\\
\hline
\end{tabular}
\begin{tabular}{l}
Two components, $\beta$ = 2, with 60 \mic \\
\end{tabular}
\begin{tabular}{llll}
 & T$_{warm}$ (K) & T$_{cold}$ (K) & $\chi^2$/d.o.f.\\
Centre & 29.0$^{+1.9}_{-3.8}$ & 15.7$^{+1.4}_{-2.4}$ & 2.88/6 \\
N-E & 29.0$^{+0.0}_{-6.1}$ & 15.7$^{+8.3}_{-0.0}$ & 5.06/6\\
S-W & 29.0$^{+1.4}_{-3.5}$ & 15.7$^{+1.3}_{-1.0}$ & 14.0/6\\
\hline
\end{tabular}
\end{center}
\end{table}

The derived parameters (temperatures and spectral indices) are consistent for the three fitted positions (centre, north-east and south-west).
However, the central peak exhibits a slightly higher temperature than the wings.
This is also true for the two-component fits and the fits with the 60 \mic~data.
The one-component fits with $\beta$ free and without the 60 \mic~data show that the wings may have a spectral index slightly lower than 2, while the centre exhibits a low spectral index (1.4).
The results show an anticorrelation between the temperature and the spectral index, which is consistent with previous results obtained by Dupac \etal (2001), Dupac \etal (2002) and Dupac \etal (2003).
However, if one considers $\beta$=2 to be the rule, then the central peak still exhibits a higher temperature (19 K) than the wings (18 K).
This is less than what derived by Israel \etal (1999) under the same assumption of $\beta$=2, with IRAS 100 \mic~and SCUBA data only (21 K).
Fitting two dust components to the emission longwards of 100 \mic~does not give very significant results: the warm temperature is not much changed with respect to the one-component fit ($\beta$=2) temperature, and the cold component is indeed extremely cold (5-9 K).
Therefore, and given the goodnesses of fit presented in Table \ref{fits}, we consider that the dust emission longwards of 100 \mic~is very adequatly fitted by a single dust population.

The fits including the 60 \mic~emission show that the one-component modeling is still possible, especially toward the centre where the $\chi^2$/d.o.f. value is low for the fit with $\beta$ free.
In this case, the derived spectral indices are low: 1.15 toward the central peak and around 1.4 toward the wings.
However, the $\chi^2$/d.o.f. values are clearly larger than those of the fits longwards of 100 \mic.
This may indicate that a two-component model is useful for fitting with the 60 \mic~emission.
This fit gives temperatures of 29 and 16 K, as well for the wings as for the centre.
Given this quite high temperature of 29 K, it is likely that the 60 \mic~emission is dominated by big grains, which gives an {\it a posteriori} consistency to this two-component model.

\subsection{Total fluxes and the mass of the interstellar medium\label{tot}}

\begin{table}
\caption[]{\label{flu}
Instrument, wavelength (\mic), total flux density integrated over three 3$'$ beams (25 kpc) along the major axis.}
\begin{flushleft}
\begin{tabular}{lll}
\hline
IRAS/HiRes & 60 & 50.5 Jy\\
IRAS/HiRes & 100 & 126 Jy\\
ISO & 170 & 197 $\pm$ 23 Jy\\
ISO & 200 & 144 $\pm$ 17 Jy\\
PRONAOS & 200 & 161 $\pm$ 10 Jy\\
PRONAOS & 260 & 120 $\pm$ 8 Jy\\
PRONAOS & 360 & 50 $\pm$ 4 Jy\\
SCUBA & 450 & 32.0 Jy\\
PRONAOS & 580 & 19 $\pm$ 3 Jy\\
SCUBA & 850 & 4.62 Jy\\
\hline
\end{tabular}
\end{flushleft}
\end{table}

For calculating the total fluxes (presented in Table \ref{flu}), we integrate the measurements over the three 3$'$ beams along the major axis, which corresponds to 25 kpc.
For estimating the mass of the galactic interstellar medium, we follow the simple model described in Dupac \etal (2001), which uses the dust 100 \mic~opacity in the diffuse interstellar medium from D\'esert \etal (1990).
In this model, the total column density of interstellar medium is simply proportional to $\epsilon_{100}$, the coefficient being 1.67$\times$10$^{24}$ H.cm$^{-2}$.
Two mass estimations are made: one assumes a single dust component with the emissivity $\epsilon_{100}$ varying along the major axis as derived by the free-$\beta$ fits without the 60 \mic~data (see Table \ref{fits}); the other assumes two dust components with varying emissivities as derived by the two-component fits with the 60 \mic~data.
In either case, we assume a distance to NGC 891 of 9.5 Mpc \cite{vanderkruit81}.
We find a mass of the interstellar medium of 4.6$\times$10$^{9}$ M$_{\odot}$ when applying the one-component free-$\beta$ model, whereas we find 12$\times$10$^{9}$ M$_{\odot}$ when applying the two-component model (including the 60 \mic~points).
This discrepancy is due to the existence of quite large amounts of cold (15.7 K) dust in the two-component results.
The value of 4.6$\times$10$^{9}$ M$_{\odot}$ is consistent with previous estimates from Gu\'elin \etal (1993): 4$\times$10$^{9}$ M$_{\odot}$, and Israel \etal (1999): 3.9$\times$10$^{9}$ M$_{\odot}$.
We can assume from Gu\'elin \etal (1993) that the atomic hydrogen mass is 2.5$\times$10$^{9}$ \msol, which gives a molecular hydrogen amount of 2.1$\times$10$^{9}$ M$_{\odot}$ from our measurement (one-component fit).
This is in good agreement with the standard assumption of similar quantities of atomic and molecular gas in spiral galaxies.
However, if we trust the two-component estimation, the molecular to atomic gas ratio becomes 3.8, which is much larger than standard assumptions.
This amount of molecular gas (9.5$\times$10$^{9}$ M$_{\odot}$) is not in good agreement with the standard estimation from CO measurements either, using the CO-to-H$_2$ conversion factor of Strong \etal (1988): about 4.5$\times$10$^{9}$ \msol.
From this two-component modeling, we derive cold over warm component mass ratios for the three studied positions: 19 for the centre, 42 for the north-eastern data point and 34 for the south-western one.
Therefore, there is a strong trend to consider that the cold (16 K) dust amount is much larger than the warm dust amount, and that the cold to warm dust ratio is twice larger in the wings of this galaxy than in the centre.
These ratios are smaller than those derived by Alton \etal (1998) with IRAS and SCUBA data only, but larger than those derived by Israel \etal (1999) with slightly different temperatures (cold to warm dust ratio of about 10).

As our ISOPHOT and PRONAOS measurements constrain very well the peak of the spectra, we are confident in the reliability of the derived dust properties (emissivity, temperature and spectral index).
However, the different possibilities to fit (well) the spectra lead to different estimates of the mass.
If one accepts the possibility that the spectral index of the dust can be somewhat different from 2, which seems wise to us regarding recent results (Walker \etal 1990, Ristorcelli \etal 1998, Dupac \etal 2001, Dupac \etal 2002, Dupac \etal 2003, Bennett \etal 2003), it does not seem possible to better discriminate between the different fitting procedures given the present data, although the spectral coverage is relatively accurate.
Also, the $\epsilon$/N$_H$ ratio used \cite{desert90} is debatable, and the mass estimates are of course very much dependent on this value.

We compare these ISM mass estimates to the {\it dust} mass (19$\times$10$^{6}$ \msol) of NGC 891 obtained by Alton \etal (2000), assuming given values for the grain radius and density.
This is substantially less than the estimation of 50$\times$10$^{6}$ \msol from Alton \etal (1998), who used a two-component model of the submillimetre emission.
Comparing the Alton \etal (2000) dust mass to our ISM mass estimates, we obtain a gas to dust mass ratio of 240 for the one-component mass estimate and 640 for the two-component one.
The first estimate of this ratio is slightly less than that derived by Alton \etal (2000): 260, a bit closer to the Milky Way value and well in agreement with the canonical dependence on the metallicity \cite{issa90}.
If we compare the ISM mass estimate from our two-component model (12$\times$10$^{9}$ \msol) to the two-component model estimate from Alton \etal (1998): 50$\times$10$^{6}$ \msol, then the gas to dust ratio is (also) 240.
We therefore consider that this value is somewhat robust.

\section{Conclusion}

We have presented PRONAOS and ISOPHOT data of the NGC 891 spiral galaxy, together with IRAS and SCUBA measurements.
The overall submillimetre spectrum of this object is now well constrained, and two different conclusions can be derived: either one single dust component is fitted to the spectra (without the 60 \mic~data), and in this case the temperature is higher toward the centre of the galaxy than toward the wings, and the spectral index is less steep (1.4 in the centre and 1.7-2 in the wings); either two components are fitted assuming $\beta$=2, and this leads to a warm temperature of 29 K and a cold temperature of 16 K.
The interstellar medium masses derived by these two methods are quite different, but both are consistent with a gas to dust mass ratio of 240.

The good spectral coverage of the submillimetre range that we present here makes us rather confident in our determination of the dust emission properties.
However, the way to fit the data, as well as the opacity properties, lead to inevitable uncertainties in the determination of the masses.
A next step in the accurate knowledge of the continuum emission of this galaxy, as well as of other nearby spirals, would be to obtain the same spectral resolution in the same wavelength range with a better angular resolution.
This should allow to better discriminate the number of distinct components, as well as the possible variations of the spectral index,
and this should be achieved by forthcoming experiments such as the Herschel satellite with the instruments PACS \cite{poglitsch01} and SPIRE \cite{griffin01}.

\section{Acknowledgments}
We thank very much R. Tuffs for his help.
We are indebted to the French space agency Centre National d'\'Etudes Spatiales
(CNES), which supported the PRONAOS project.
We are very grateful to the
PRONAOS technical teams at CNRS and CNES, and to the NASA-NSBF balloon-launching facilities group of Fort Sumner (New Mexico).
We thank very much the technical and scientific teams of ISO, an ESA project with instruments funded by ESA member states (especially the PI countries: France, Germany, the Netherlands and the United Kingdom) with participation of ISAS and NASA.

\end{document}